\documentclass[screen]{acmart}
\usepackage{graphicx}
\newcommand{\fixme}[1]{{\color{black}#1}}
\AtBeginDocument{%
  \providecommand\BibTeX{{%
    \normalfont B\kern-0.5em{\scshape i\kern-0.25em b}\kern-0.8em\TeX}}}

\setcopyright{acmlicensed}
\copyrightyear{2018}
\acmYear{2018}
\acmDOI{XXXXXXX.XXXXXXX}

\acmConference[Conference acronym 'XX]{Make sure to enter the correct
  conference title from your rights confirmation emai}{June 03--05,
  2018}{Woodstock, NY}
\acmISBN{978-1-4503-XXXX-X/18/06}
\usepackage{tabularx}
\usepackage{enumitem}
\usepackage{svg}
\usepackage{rotating}
\usepackage{adjustbox}
\usepackage{array}
\usepackage{amsmath} 
\usepackage{nameref}
\usepackage{hyperref}

\begin{document}

\title[Exploring Viewing Modalities in Cinematic Virtual Reality]{Exploring Viewing Modalities in Cinematic Virtual Reality: A Systematic Review and Meta-Analysis of Challenges in Evaluating User Experience}


\author{Yawen Zhang}
\email{mozwen17@gmail.com}
\orcid{}
\affiliation{%
  \institution{City University of Hong Kong}
  \streetaddress{}
  \city{Hong Kong}
  \state{}
  \country{China}
  \postcode{}
}

\author{Han Zhou}
\email{zh.zhouhan.z.h@gmail.com}
\affiliation{%
  \institution{University of Wisconsin-Madison}
  \streetaddress{}
  \city{Madison}
  \country{USA}}

\author{Zhongmingju Jiang}
\email{}
\affiliation{%
  \institution{Southern University of Science and Technology}
  \city{Shenzhen}
  \country{Chian}
}

\author{Zilu Tang}
\email{}
\affiliation{%
 \institution{Eindhoven University of Technology}
 \city{Eindhoven}
 \country{Netherland}
 }
\author{Tao Luo}
\authornote{Corresponding author}
\email{luot@sustech.edu.cn}
\affiliation{%
  \institution{Southern University of Science and Technology}
  \streetaddress{}
  \city{Shenzhen}
  \country{China}}
  
\author{Qinyuan Lei}
\email{qinyulei@cityu.edu.hk}
\authornote{Corresponding author}
\affiliation{%
  \institution{City University of Hong Kong}
  \streetaddress{}
  \city{Hong Kong}
  \country{China}}


\begin{abstract}
 Cinematic Virtual Reality (CVR) is a narrative-driven VR experience that uses head-mounted displays with a 360-degree field of view. Previous research has explored different viewing modalities to enhance viewers’ CVR experience. This study conducted a systematic review and meta-analysis focusing on how different viewing modalities, including intervened rotation, avatar assistance, guidance cues, and perspective shifting, influence the CVR experience. The study has screened 3444 papers (between 01/01/2013 and 17/06/2023) and selected 45 for systematic review, 13 of which also for meta-analysis. We conducted separate random-effects meta-analysis and applied Robust Variance Estimation to examine CVR viewing modalities and user experience outcomes. Evidence from experiments was synthesized as differences between standardized mean differences (SMDs) of user experience of control group (“Swivel-Chair” CVR) and experiment groups. To our surprise, we found inconsistencies in the effect sizes across different studies, even with the same viewing modalities. Moreover, in these studies, terms such as “presence,” “immersion,” and “narrative engagement” were often used interchangeably. Their irregular use of questionnaires, overreliance on self-developed questionnaires, and incomplete data reporting may have led to unrigorous evaluations of CVR experiences. This study contributes to Human-Computer Interaction (HCI) research by identifying gaps in CVR research, emphasizing the need for standardization of terminologies and methodologies to enhance the reliability and comparability of future CVR research.
\end{abstract}



\maketitle

\section{Introduction}
\label{sec:introduction}
\indent With the development of virtual reality (VR) technology, traditional media has undergone a revolution, achieving transformation through the interactive capabilities and technical immersion provided by VR \cite{bates1992virtual,biocca2013communication}. Cinematic Virtual Reality (CVR) refers to the experience of using a VR device, especially a head-mounted display (HMD), as a medium for watching narrative-based 360-degree videos \cite{macquarrie2017cinematic,mateer2017directing,rothe2019guidance}.

Unlike traditional films, CVR is “frameless,” offering viewers a 360-degree field of view (FOV) \cite{bucher2017storytelling,dooley2017storytelling}. Many studies have aimed to establish a new “grammar” for CVR narratives through this frameless feature \cite{VR2016brillhart,doyle2023grammar,lescop2017narrative} from both technical and narrative perspectives \cite{zhang2020developing,aylett2003towards}. However, existing studies have often failed to control both aspects, which shows that the development of CVR’s grammar is still in its early stages \cite{rothe2017diegetic,mateer2017directing,agullo2019making}.

During the development of the new grammar, researchers conducted various experiments on viewing modalities to enhance viewers’ CVR experience. More specifically, modalities optimized through technological means can be divided into two categories—namely, intervened rotation \cite{rothe2020reduce,aitamurto2021fomo} and avatar assistance \cite{dining2017user,masia2021influence}. Other modalities aim to improve narrative languages through guidance cues (diegetic/non-diegetic; \cite{nielsen2016missing,rothe2019guidance,cao2020automatic}), perspective shifting (first-person perspective (1PP)/third-person perspective (3PP); \cite{bender2019headset,bahng2020reflexive,cannavo2023immersive}), and other novel editing methods \cite{VR2016brillhart,sassatelli2018snap}. 

However, there is a notable lack of systematic analysis comparing the outcomes of different viewing modalities on CVR experience. This lack of systematic analysis in turn hinders future studies in CVR in Human-Computer Interaction (HCI), more specifically, in CSCW topics such as social viewing in CVR \cite{Siri2018Screen, Zamanifard2019Togetherness, Kukka2017SocialCity, montagud2022towards}, collaborative AR/VR in work environments \cite{kohler2011avatar,Ortiz2023Workspace}, and collaborative storytelling in CVR. This study aims to address this research gap by conducting a systematic and statistical review of viewing modalities, summarizing the characteristics and the effects of different viewing modalities on CVR experience, paving the way for further research. This study, therefore, starts with a meta-analysis to answer the following research question:

\emph{\textbf{RQ1: Based on the evidence of previous studies, how do different viewing modalities affect CVR experience?}}

To evaluate the effectiveness of viewing modalities, researchers typically employ subjective measurements, especially questionnaires. These questionnaires, when used in CVR empirical research, often adopt metrics originally developed for VR research. These metrics primarily measure users’ subjective experiences in VR environments, such as their sense of presence \cite{igroup,usoh2000using,witmer1998measuring}, sense of immersion \cite{jennett2008measuring}, and motion sickness \cite{kolasinski1995simulator}. Additionally, CVR research also incorporates metrics from traditional cinema, focusing on viewers’ subjective responses to changes in narrative language. Other metrics commonly used by researchers include narrative engagement \cite{busselle2009measuring}, empathy \cite{raes2011construction}, and happiness \cite{ip2018design}. The numerous different metrics used to evaluate the CVR experience makes it difficult to create guidelines in user studies. Therefore, to help HCI researchers determine clear guidelines for user studies in CVR, we propose two additional research questions following up on RQ1:

\emph{\textbf{RQ2: What are the primary methods and key metrics currently used for measuring user experience in CVR?}}

\emph{\textbf{RQ3: What are the current issues with evaluation and metrics in measuring different viewing modalities?}}

In this study, we conduct a systematic review to gain an overview of the CVR field as well as a meta-analysis to collect and analyze quantitative data. First, we gather existing research on CVR from several databases, amounting to 3444 papers. Second, we conduct literature screening and exclusion following the PRISMA guidelines \cite{tricco2018prisma}, resulting in 45 papers for the systematic review, 13 of which are also used for the meta-analysis. The systematic review focuses on investigating research topics and the status of existing evaluation methods used in CVR studies. In the meta-analysis, we focus on clarifying how studies have explored different viewing modalities and how these affect viewer experience. \fixme{We found that certain viewing modalities have potential impacts on the user experience: Modality 2 and Modality 5 show potential positive impacts, while Modality 6 might lead to negative outcomes, echoing findings from earlier studies (see~\ref{subsec:4.2Meta}).}

However, one of our most surprising findings is that there are inconsistencies in the effect sizes across different studies, even with the same viewing modalities, which could be attributed to internal differences within the experimental designs. Moreover, we find that researchers predominantly use questionnaires that mainly focus on “presence” or “immersion.” These terms, along with “narrative engagement,” are often used interchangeably. Overall, our findings show that the irregular use of questionnaires, overreliance on self-developed questionnaires, and incomplete data reporting may have led to unrigorous evaluations of CVR experiences in user studies.

\section{Related Work}

\subsection{What is Cinematic Virtual Reality?}

Cinematic Virtual Reality (CVR) is defined as an evolving genre of immersive experiences, where individual users engage with synthetic worlds in 360° \cite{mateer2017directing}. The viewer watches stereoscopic movies with spatialized audio meticulously designed to enhance the authenticity of the virtual environment \cite{tong2022adaptive,rothe2019guidance}. CVR is also defined by some scholars as a term that encompasses a broad range of content, from passive observation of omnidirectional movies to more interactive narrative-driven experiences \cite{macquarrie2017cinematic,szita2018effects}. Some scholars also use the term “Film VR”, “VR film” or “VR movie” \cite{syrett2017oculus,ross2018cinematic,serrano2017movie}. We adopted a broad definition of Cinematic Virtual Reality, that CVR is the experience of using Virtual Reality (VR) devices, especially Head-Mounted Displays (HMD), as the medium to watch narrative-based 360° videos. By focusing on the use of VR devices, particularly HMDs, we acknowledge the pivotal role of technology in shaping the immersive qualities of CVR experiences. The adoption of this definition also recognizes the transformative impact of VR on traditional cinematic storytelling, where users are not merely spectators but active participants in the narrative \cite{ryan2004will,mateas2000neo} with the freedom to control their field of view and engage with the content in a more interactive and personalized manner \cite{macquarrie2017cinematic,dooley2017storytelling}.

Media content in VR can be classified into different categories based on their level of interactivity \cite{tong2021viewer}. CVR typically promotes a “lean back” interactivity style \cite{vosmeer2014interactive}, which encourages more passivity in the user and allows the user to observe the story unfold \cite{rothe2019guidance}. The most typical type of CVR is called “Swivel-Chair” CVR [53], in which viewers’ positional movements and interactions with other characters or elements in the virtual environment are limited providing less interactivity. Another type of CVR called interactive CVR allows viewers to have more interactivities by using some interactive methods, such as hand controllers \cite{beck2021applying,tong2022adaptive} and eye-tracking \cite{Drewes2021Gaze}. 
 
For CVR content creators, finding the right balance between interactivity and narrative immersion is a significant challenge due to the “Narrative Paradox” \cite{aylett2000emergent,roth2018ludonarrative}. Narrative Paradox points to the phenomenon that the two key aspects of CVR, namely interactivity and narrative immersion, often appear to be at odds with each other. Increased interactivity in VR content may be achieved at the expense of the immersive storytelling experience, and vice versa. The narrative core of CVR is what sets it apart from other 360° videos and other VR media. The emergence of CVR has prompted a reevaluation of traditional storytelling tools and techniques \cite{serrano2017movie, VR2016brillhart}. This presents storytellers with the challenge of effectively guiding users while conveying a narrative coherently \cite{cheever2014out}.

CVR is not only used for entertainment, but also for various purposes, such as education \cite{zhang2019exploring,queiroz2023collaborative}, training \cite{love2023uses}, social support \cite{rawski2022sexual}, and historical tours \cite{argyriou2020design,franccois2021virtual}.  Due to CVR's narrative core, it is effective in conveying stories uniquely and evoking empathy by immersing users in highly affective spaces \cite{bollmer2017empathy}. Notably, a lot of research uses a wide range of techniques and tools to adjust the viewing modalities to leverage the unique immersive characteristics of VR, making it a truly immersive medium for storytelling. Our focus in this paper is to investigate the different viewing modalities of CVR by examining both their benefits and drawbacks.

Previous studies have partially reviewed various aspects of CVR as an emerging field. For instance, Rothe et al.\cite{rothe2019guidance} and Dooley \cite{dooley2021cinematic} delved into the mechanisms of user attention in CVR, proposing a detailed taxonomy of guidance strategies \cite{rothe2019guidance,dooley2021cinematic}. Additionally, a recent review by Yu \& Lo \cite{yu2023mapping} highlights the impact of VR headset, spatialized audio, and interactivity effectiveness of VR headset iteration, spatialized audio, and interactivity, on the immersive experience in CVR \cite{yu2023mapping}. However, these studies only focus on selected aspects of CVR and often lack methodological comprehensiveness. We aim to fill this research gap by systematically categorizing and defining the viewing modalities in CVR, offering a holistic perspective that encompasses the diverse ways users consume CVR content. This comprehensive approach aims to provide a foundation for evaluating user experience in CVR, addressing the multifaceted nature of this emerging field.
\subsection{Viewing Modalities in CVR}
In previous CVR studies, researchers developed new viewing modalities in the following four directions: guidance cues, intervened rotation, perspective shifting, and avatar assistance \cite{martin2022multimodality}. Since attention is a crucial factor in both CVR, effectively managing it is essential for creating immersive and engaging experiences. Consequently, guidance cues are applied in CVR to direct the viewer’s attention and convey narrative information \cite{veas2011directing,rothe2019guidance,sheikh2016directing}, enhancing immersion and preventing the fear of missing out (FOMO) \cite{cheever2014out}. Nielsen et al. \cite{nielsen2016missing} first classified guidance cues in CVR into three orthogonal, dichotomous dimensions. In their taxonomy, the most recognized category is diegetic or non-diegetic, based on whether the cues exist within the fictional world of the narrative or outside it \cite{nielsen2016missing,rothe2019guidance}. To guide the viewer’s attention, creators can use diegetic cues that are included in the story world, such as moving protagonists \cite{xu2019effects,cao2020automatic}, lights, or sounds \cite{nielsen2016missing,beck2021applying,pillai2019grammar}, while there are also non-diegetic cues that are not part of the story, such as arrows and flickers \cite{cao2020automatic,rothe2018guiding,speicher2019exploring}. The effective use of either type of cues, or sometimes a combination, allows creators to strike a balance between immersion and control, ensuring that viewers are engaged with the content while also having the freedom to explore and make choices within the narrative context \cite{rothe2019guidance,mateer2017directing}.

Controlling the viewer’s freedom of rotation in CVR is another technique used to influence the viewer’s attention and guide their experience. This approach can be valuable for creators who want to ensure that viewers focus on specific elements or storytelling aspects within the virtual space, such as FOMO in the movie \cite{aitamurto2021fomo}. There are two primary ways to control the viewer’s rotation, which are assisted and limited. The choice between rotation restriction and enhancement depends on the goals of the VR experience and the desired level of user agency. Rotation-assisted techniques are more suitable for experiences that encourage exploration. Hong et al. \cite{hong2016accelerated} amplified the rotational angle with a 1.3–1.6 scale. Rothe et al. \cite{rothe2020reduce} utilized a controller to help users rotate their views at different speeds. Rotation limitation is often used when precise control over the viewer’s perspective is necessary for storytelling or instructional purposes. Aitamurto et al. \cite{hong2016accelerated} tested the half-sphere condition to compare its effects on FOMO. Rothe et al. \cite{rothe2020reduce} compared three rotational methods with 360, 180, and 225 degrees.

Shifting perspectives within the same CVR scenario is another valuable approach used in guiding the viewing modalities of the user \cite{gorisse2017first}. These shifts can occur when users view the same scene or narrative event from multiple vantage points. They offer insights into how viewers engage with the content and how their perception and emotional responses may vary \cite{bahng2020reflexive,cao2019preliminary,bender2019headset}.

As a viewing modality, avatars can also play a crucial role in shaping the narrative, engagement, and emotional connection within the virtual environment \cite{schultze2010embodiment,gunther2015aughanded}. The movements and actions of the viewer are reflected in real-time by their avatar in the virtual environment, creating a sense of physical presence and immersion \cite{christy2016transportability}. Moreover, avatars provide more possibilities for transportability, which is an important factor in increasing identification with story characters \cite{chen2017effect,dal2007smoking,kohler2011avatar}. The absence of the virtual body sometimes has positive effects, since it provides moments of reflection and relaxation during which viewers can immerse themselves in a serene or meditative environment \cite{bahng2020reflexive}. The presence or absence of an avatar is a creative decision that shapes the narrative and emotional impact of VR experiences, making avatars versatile in and adaptable to different types of storytelling.

\subsection{Evaluation of CVR Viewing Modalities}
Evaluation of the CVR experience is crucial for testing the benefits of different viewing modalities. In more than a decade of studies \cite{veas2011directing}, previous work gradually completed and refined the metrics that CVR research is generally interested in—namely presence, immersion, motion sickness, memory, and level of enjoyment \cite{mutterlein2018three}.

Considering the inherent characteristics of the VR system, presence and “immersion, the most common metrics in the CVR experience, are what many VR researchers consider be important \cite{schwind2019using,lombard1997heart,mestre2006immersion}. The sense of presence denotes the viewer’s psychological perception of “being there” or “existing in” a virtual environment \cite{heater1992being,slater2003note}. Slater \cite{slater2003note} defined immersion as “the objective level of fidelity of the sensory stimuli produced by a VR system.” Unlike presence, the immersion level of a VR system depends more on technical aspects such as software rendering quality or hardware display technologies \cite{elmezeny2018immersive,rothe2019guidance}. However, the definitions of these terms are still ambiguous \cite{nilsson2016immersion,ermi2005fundamental,grassini2020questionnaire}. There is an overlap in these definitions of these terms, with all describing viewers’ subjective evaluation of their own engagement. McMahan \cite{mcmahan2013immersion} criticized immersion as becoming “an excessively vague, all-inclusive concept” \cite{sheikh2016directing}. In practical usage, there is an inconsistency between presence and immersion, and the terms are sometimes even used interchangeably with concepts such as narrative engagement, user engagement, and involvement \cite{nilsson2016immersion,macquarrie2017cinematic,ryan2015narrative}.

Motion sickness, also known as simulator sickness \cite{kolasinski1995simulator}, VR Sickness \cite{chang2020virtual}, or cybersickness \cite{dennison2016use,davis2014systematic}, is another necessary result measured in previous research. Motion sickness is caused by the visual–vestibular conflict, a sensory mismatch between visual and vestibular information regarding motion and spatial orientation \cite{akiduki2003visual,treisman1977motion,cheever2014out}, resulting in symptoms such as dizziness, eyestrain, and headache \cite{kennedy1993simulator}, which can evidently impair the user experience \cite{chang2020virtual,nielsen2016missing}. The simulator sickness questionnaire (SSQ) \cite{kennedy1993simulator} is commonly used to measure motion sickness or cybersickness in HCI applications. However, While the SSQ has a wide range of applications, it was not developed with VR scenarios in mind; its scope extends beyond VR-specific contexts. Recently, many other questionnaires are also commonly used in different areas related to VR \cite{jeong2023mac,tasnim2024investigating,islam2021cybersickness}, such as VRSQ \cite{KIM201866} and fast-motion scale questionnaires (FMSQ) \cite{FMSQ2011Behrang}.

Moreover, memory is always measured after the CVR experience. This can be done by asking stimulus content-related questions using the Wechsler memory scale \cite{wechsler1945wechsler}. Free recall \cite{roediger2006power} and cued recall \cite{furman2007they} are two optional ways to collect users’ memory data. Although there is not sufficient evidence from previous research to prove the relationship between recall performance and immersion, some researchers believe that better memorization presents higher immersion \cite{ragan2010effects}. However, studies have found that user memory can be easily affected by many other factors, such as different emotions \cite{chirico2017effectiveness,allcoat2018learning} or the FOMO \cite{alt2015college,aitamurto2021fomo}. 

Level of enjoyment is a straightforward measurement used in some studies. Many researchers are convinced that \cite{xu2019effects} the level of enjoyment can reflect the degree of user engagement \cite{tong2022adaptive,vosmeer2017you}. However, due to the current lack of a valid measurement for enjoyment, most studies have resorted to using simple self-developed questions. In summary, there are no specialized questionnaires designed specifically to evaluate CVR experiences, which is the aim of our future research.

\section{METHODS}
\subsection{Literature Search}
For this systematic review and meta-analysis, we searched Scopus, Web of Science Core Collection, PsycINFO, IEEE Xplore and ACM Digital Library for studies published between Jan 1, 2013 and June 17, 2023. The following search terms were used: “Virtual Reality” and “VR,” combined with “cinematic,” “cinema,” “movie,” and “film.” We focused on peer-reviewed articles published in English within the last 10 years (since 2013). The search yielded 3442 results. Furthermore, we searched citations from other review papers related to CVR research \cite{rothe2019guidance, dooley2021cinematic,yu2023mapping}, and found two more papers \cite{gugenheimer2016swivrchair,lin2017tell} that meet our inclusion criteria.

\begin{figure}[ht]
\centering
\includegraphics[width=0.7\linewidth]{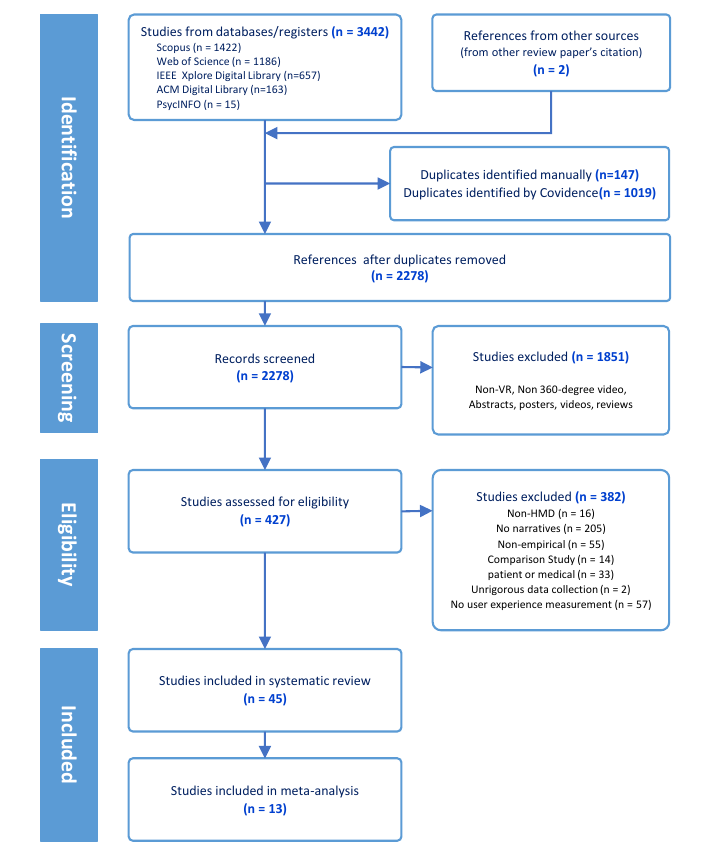}
\caption{A PRISMA-style flow diagram of the literature selection process}
\label{fig:PRISMA}
\end{figure}
\subsection{Inclusion Criteria}
To gain a better understanding of the current state of evidence-based research in the field of CVR, we selected articles for analysis that met the following criteria: studies 1) employing a survey or experimental design where participants were assigned to watching CVR (narrative 360-degree videos) with HMDs; 2) with participants acting as the viewer (this would exclude work using a manipulative virtual environment, task-based games, or any content not related to narratives); and 3) measuring user experience. \fixme{Please note that papers on social VR are not included in our corpus, because they primarily focus on social interactions and do not examine how specific viewing modalities influence user experience. For discussion of how our research will inform future studies on social VR, please see Section \ref{subsec:5.3ResearchImplications}.} Reports on quantitative statistics (e.g., correlation coefficient, regression coefficient) were kept for meta-analysis. The PRISMA flowchart in Figure. \ref{fig:PRISMA} shows the details.

\subsection{Screening Procedure}
First, we removed duplicates by using an automated tool, Covidence \cite{babineau2014product}, and manual identification, leaving 2278 records. Second, three researchers conducted a title and abstract screen in Covidence on the remaining studies. During this phase, each study’s title, abstract, and publication information exported from the databases was manually screened and checked by at least two of the three reviewers. Of these, 1851 were excluded as irrelevant, and 427 studies were identified for further checking. Finally, after full-text screening (each study was screened by two coders, see the details of coding procedure in~\ref{subsec:3.4coding}), we reviewed all the selected articles in detail. 45 articles that meet the inclusion criteria were finally included (13 papers were selected for meta-analysis).

\subsection{Coding Procedure}
\label{subsec:3.4coding}
We implemented a systematic process to identify and code the features of each study included in our systematic review and meta-analysis. For each paper that met the meta-analysis eligibility criteria, we manually extracted relevant variables and collected the information shown in Table \ref{tab:code}: 

\begin{table}[htbp]
\centering
\caption{Coding book}
\label{tab:code}
\begin{tabular}{@{}p{4cm}p{11cm}@{}}
\toprule
\textbf{Category} & \textbf{Details} \\
\midrule
\textbf{Basic Information} & Study ID, Effect Size ID, Author, Publication Year \\
\addlinespace[10pt]
\textbf{Research Topics} & Topics such as guidance, editing, presence, as well as trends \\
\addlinespace[10pt]
\textbf{Study Design and \newline Participants} & Sample size; Study design (whether it is within-subjects); \newline Sampling type (random or not); \newline Use of controller (with or without); \newline Viewing modalities (with explicit diegetic guidance, with explicit non-diegetic guidance, with implicit non-diegetic guidance, with agency, limited rotation, forced rotation); \newline Storytelling (linear or non-linear); Spatial sound (with or without); \newline Perspective (First Person Perspective or not); Stimulus length; \newline Format (real-time rendering VE, film); Context (narrative, documentary, scenes);  \\
\addlinespace[10pt]
\textbf{Measurements} &  Type of questionnaire (adapted, valid, selected items, self-developed); \newline Use of terminology (whether it contains mixed terms); \newline Objective measurements (e.g., eye-tracking technology) \\
\addlinespace[10pt]
\textbf{Statistical Results} &  effect size (Hedge’s g) was computed using means and standard deviations from groups with different viewing modalities (the control group is “Swivel-Chair” CVR), or reports of F-test result, t-value, p-value  \\
\bottomrule
\end{tabular}
\end{table}

The coding process was conducted by the first and third authors. Initially, the two coders discussed the codes together to ensure clarity and mutual understanding of all items. Subsequently, 20\% of all articles were randomly selected as a sample and distributed to both coders. The coding was performed independently by each coder. Upon finishing, the two coders compared and discussed their results. The inter-rater agreement rate reached 87.5\% [i.e., $\frac{8-1}{8}=87.5\%$], indicating high consistency between the two coders. Based on the preliminary coding results, the research team discussed issues revealed in the initial coding process. A revised coding plan was developed, which involved adding and deleting certain coding features. Finally, all 45 articles were carefully coded by both coders using this revised coding plan.

\subsection{Analytic Procedures of the Meta-Analysis}
This meta-analysis aims to examine the effects of different viewing modalities on user experience in CVR including presence, immersion and narrative engagement. Effect sizes were operationalized as standardized mean differences (SMDs) of user experience between the control group (“Swivel-Chair” CVR) and the experiment groups. These were coded so that positive values indicated better outcomes (e.g., more narrative engagement). However, when conducting the meta-analysis, we noticed that the extracted effect sizes were not independent, as one study might have reported multiple effect sizes and one control group was used with multiple treatments. To account for this dependence structure, we employed two analytical approaches. The first approach was to create sub-sets of effect sizes based on the six categories of viewing modality, where each sample has at most one effect size estimate per sub-set. However, the limited effect sizes within each group (ranging from 2 to 6) presented significant challenges for moderator analysis. Consequently, we decided to take the second approach by conducting a separate meta-analysis for each group. We used the correlated hierarchical effects (CHE) working model with robust variance estimation, as proposed by Pustejovsky and Tipton (2022) \cite{pustejovsky2022meta}. This method examined a range of moderator variables related to viewing conditions and measurement characteristics, while also providing descriptive information about both the between- and the within-study heterogeneity. Due to the limited information about the sampling correlations among effect sizes, we assumed varying constant sampling correlations of $\rho$ = 0, 0.3, 0.6 and 0.9. Data analysis was conducted using the \texttt{metafor} package, the \texttt{robumeta} package and the \texttt{clubSandwich} package \cite{viechtbauer2010conducting} in R.\\

\section{RESULTS}
\subsection{Preliminary View of Systematic Review}
This section provides an overview of the systematic review, offering a preliminary analysis of the field, including research topics and trends; publication information; studies, apparatus and stimulus; and participants.

\subsubsection{Research Topics and Trends}

In this section, we present the results of our review. First, we provide an overview of the articles’ research topics (seven research topics were selected, coded from a to g, as in the following descriptions) and publication origins. Then, we analyze and discuss the trends of CVR research in the field. 

\begin{figure}[h]
 \includegraphics[width=0.8\linewidth]{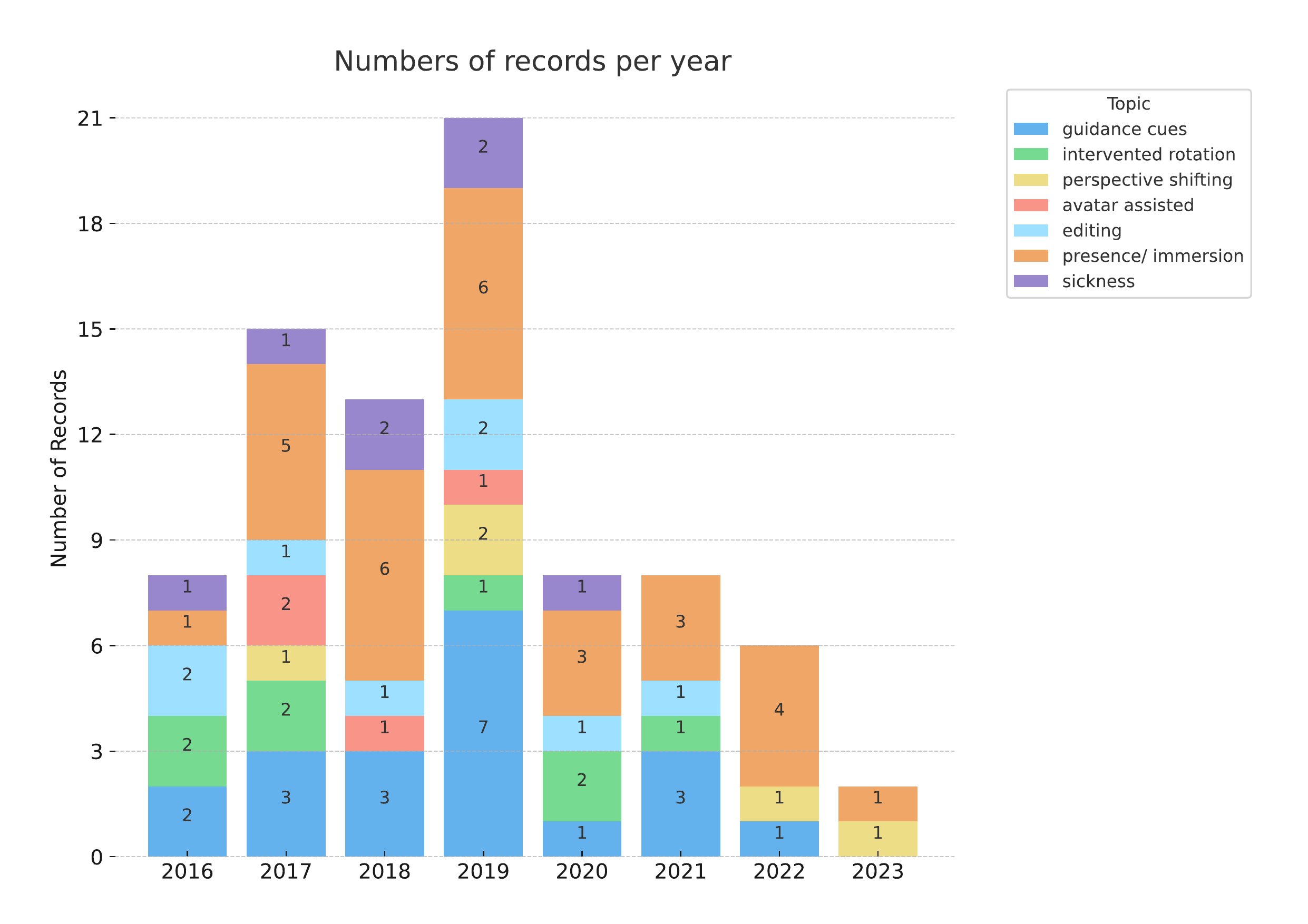}
 \caption{Number of records for different research topics per year}
 \label{fig:Fig2}
\end{figure}

\begin{enumerate}[label=(\alph*)]
\item \emph{Guidance cues}: methods and techniques for guiding the audience’s attention, including diegetic and non-diegetic guidance or visual and sound guidance.
\item \emph{Intervened rotation}: methods for manipulating the user’s rotational movement, such as forced rotation and assisted rotation.
\item \emph{Perspective shifting}: techniques for changing the user’s viewing perspective, such as a first-person or a third-person perspective.
\item \emph{Avatar assistance}: methods for setting a virtual body to map the movements and actions of the viewer in real-time in the virtual environment, creating a sense of physical presence and immersion.
\item \emph{Editing}: techniques for creating a sense of situational continuity, such as fade-out, fade-in, and dissolve. For VR content editing, the type of edit includes the cognitive point of view of event segmentation, the number and position of regions of interest before and after the cut, and their relative alignment across the cut boundary of situational continuity.
\item \emph{Presence and immersion}: research on the viewers’ psychological perception of being there or existing in a virtual environment and the objective level of fidelity of the sensory stimuli produced by a VR system.
\item \emph{Sickness}: research on the level of motion sickness caused by different CVR content or viewing modalities.
\end{enumerate}

Research on CVR began to emerge in 2016 and peaked in 2019 (See Figure \ref{fig:Fig2}). However, there has been a significant decrease since 2019, likely due to limitations on empirical research brought about by the COVID-19 pandemic. In the published studies in 2019, there was a notable emphasis on guidance cues as well as presence and immersion. Through our literature review, we noticed that articles centered on guidance cues frequently discussed the impact of cues on presence or immersion.

Over the years, the research themes of guidance cues, presence and immersion, and editing have consistently remained popular. Despite fluctuations in the number of published studies, these themes have consistently been mentioned over the years. On the other hand, research themes such as intervened rotation and avatar assistance are featured few papers, suggesting their potential niche focus within the field. Meanwhile, sickness and perspective shifting made occasional appearances in the data, indicating their limited representation in certain years.

\subsubsection{Publication Information}
In terms of the origins of these papers, the compilation consisted of 41 conference papers and four journal articles. Among the conference papers, distribution across various scholarly platforms became evident. These interdisciplinary conferences gathered the latest academic research findings related to CVR. They included authoritative conferences in the HCI field, such as CHI (seven papers, 15.6\%); HCII (three papers, 6.7\%); VR-related conferences, such as IEEE VR (5 papers, 11.1\%); and conferences emphasizing interactive narratives, such as the International Conference on Interactive Digital Storytelling (ICIDS, three papers, 6.7\%).

\subsubsection{Studies, Apparatus and Stimulus}
Among the 51 experiments conducted across the 45 papers (with a median of one experiment and SD=0.40), 33 (64.7\%) studies had within-subjects designs and 17 (33.3\%) were between-subjects experiments. One paper lacked detailed reporting. 

Various HMD devices were used in these experiments, including HTC Vive (n=12), Samsung Gear VR (n=13), Oculus Rift (n=6), and other versions of HMD devices, such as Oculus Quest 1/2, Oculus Go, Fove-DK-0, iQUT, and Daydream. Additionally, a diverse array of auxiliary devices were employed, including eye-tracking devices, smartphones (e.g., Samsung Galaxy 6), Bluetooth controllers, and computers.

In terms of experiment stimulus, there were the following three main types of stimulus: films (n=31, 60.7\%), animated videos (n=13, 25.5\%), and real-time rendering scenes (n=7, 13.7\%). As for the stimulus content, among the subset of films and animated videos, 34 were narratives and 10 documentaries. We observed no discernible pattern in the reported durations of the stimuli. The mean length was 9 minutes and 5 seconds, but the range varied significantly (from 1 minute to 30 minutes).
Regarding perspectives, the results indicated a balanced trend between the use of the 1PP and the 3PP (n=25, n=22, respectively). Furthermore, four papers employed a combination of both perspectives or incorporated shifting perspectives.

\subsubsection{Participants}
A cumulative total of 1872 individuals (M=38.2, SD=30.8) took part across 49 data experiments. Two outliers-n=3259 \cite{maranes2020exploring} and n=317 \cite{gospodarek2019sound}-were excluded from statistics. There were three experiments that omitted the reporting of participant numbers. Although the range of participants varied notably (n=12-165), most of the papers included a relatively limited number of participants (n$\le$45). 

Our results showed no gender distribution tendency. There were 799 male and 677 female participants in 41 experiments (of the 51 experiments in total, only 41 reported on gender), representing approximately 54.1\% and 45.9\%, respectively. Furthermore, we found that the age range of participants was remarkably diverse, ranging from 12 to 76 years.

\subsection{Meta-Analysis: Variability in Results and Effect Sizes Across Experiments}
\label{subsec:4.2Meta}
Thirteen articles in the meta-analysis were published between 2016 and 2022, including 14 studies and 26 effect sizes. The samples in these studies ranged from 12 to 44 in size. Detailed information on these studies can be found in the Appendix (See Figure \ref{fig:appendix2}), as well as further detailed information on their use of questionnaires and terminologies (See Table \ref{tab:Q&T}).

\emph{Publication Bias. } We proceeded to Egger’s test \cite{egger1997bias}, with the results showing a p-value of >0.05 (z=-0.97, p=0.33), suggesting that publication bias could be ignored \cite{hunter2004methods}, a symmetric distribution of the funnel plot is depicted in the Appendix (See Figure \ref{fig:appendix1}).

\begin{figure}[ht]
\centering
 \includegraphics[width=1.1\linewidth]{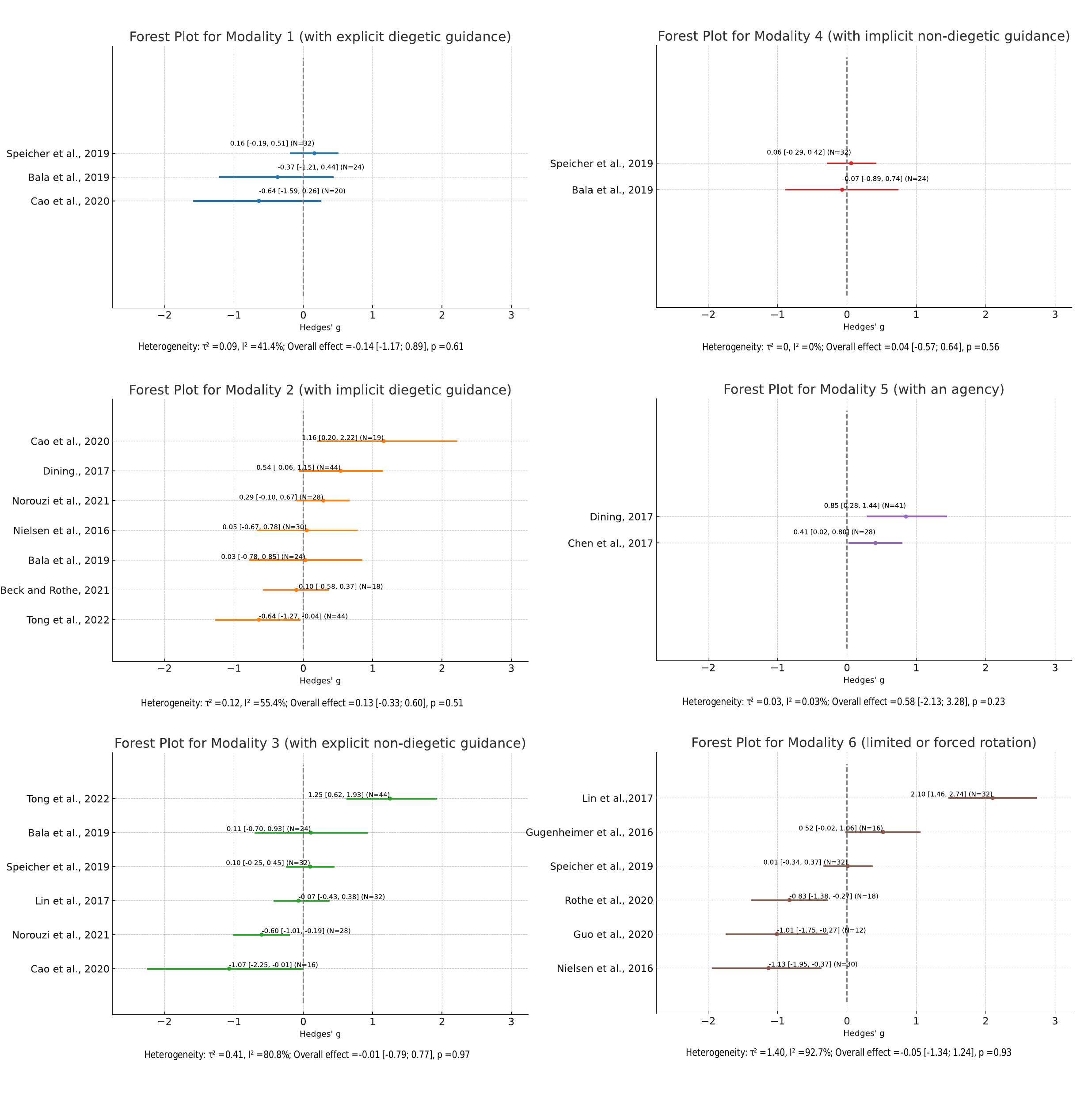}
  \caption{Forest plot}
  \label{fig:forest plot}
\end{figure}

\emph{Overall Effect Size. } The effect size of the included studies on viewing modality and the presence/immersion/engagement reports (See Figure \ref{fig:forest plot}) varied greatly, from g=-1.13 \cite{cannavo2023immersive} to g=2.10 \cite{lin2017tell}. The Hedge’s g-value was negative, indicating that the viewing modality can reduce the presence/immersion/engagement effect. Without considering the dependence of these effect sizes, the overall effect was Hedge’s g=0.070, SD=0.127. It had a very small effect size and was not significant (t(25)=0.488, p=.63 > .001), with a confidence interval between -0.225 and 0.365.

The following results of meta-analysis show the answer 
 of \noindent\emph{\textbf{"RQ1: Based on the evidence of previous studies, how do different viewing modalities affect CVR experience?"}} 
 
 \fixme{Overall, we did not observe significant effect sizes in any modality, although some modalities show potential impacts. For viewing modalities that provide visual guidance cues (Modality 1-4), results can vary dramatically depending on the setting: implicit and diegetic guidance tend to have a positive effect, with implicit diegetic guidance (Modality 2) most likely enhancing the viewing experience. In contrast, viewers may find explicit and non-diegetic guidance abrupt or disorienting. This aligns with the findings of Rothe et al. (2019) on guidance taxonomy in CVR \cite{rothe2019guidance}. Moreover, while underexplored, Modality 5 shows moderate positive effects, supporting previous research that avatars can enhance presence in VR environments \cite{Mottelson2023, Freeman2021}. Our results suggest that avatars could potentially lead to positive impacts in CVR. Limited or forced rotation (Modality 6) generally shows negative results due to increased motion sickness, unless specific design settings are implemented to reduce this effect, which is consistent with previous research \cite{chang2020virtual, weech2019presence}}
\\

\subsubsection{The results of sub-sets analysis}
\emph{\textbf{Modality 1 (with explicit diegetic guidance): }}Both positive and negative results were present. The sample sizes were generally small. There were also considerable variations in the design of the experiments.
\newline \emph{\textbf{Modality 2 (with implicit diegetic guidance): }}Overall, two effect sizes in two studies \cite{beck2021applying,tong2022adaptive} were negative, while the other results were slightly or significantly positive. One possible cause of the diminishing immersion/presence is that the two studies both had controllers, which allowed the viewers to have more interactive freedom. 
\newline \emph{\textbf{Modality 3 (with explicit non-diegetic guidance): }} There was significant diversity in the data but no clear patterns. Due to the varying interpretations of “explicit” visual guidance cues among researchers, there was significant disparity in the design of stimuli. For example, the experiment by Tong et al. \cite{tong2022adaptive} utilized wireframes closely aligned with the scene setting as guidance, while Cao et al. \cite{cao2020automatic} designed their guidance using highly conspicuous patterns in a dark virtual environment.
\newline \emph{\textbf{Modality 4 (with implicit non-diegetic guidance): }} The overall effect size showed a slight positive effect, but the CI included zero and low heterogeneity. The result was not statistically significant. Both studies had a small sample size.
\newline \emph{\textbf{Modality 5 (with an agency): }} This viewing modality was not adequately explored. However, a moderate positive effect was observed in the current studies, suggesting the significant effect of an avatar in CVR, which may enhance presence. 
\newline \emph{\textbf{Modality 6 (limited or forced rotation):  }} Most results in this modality were negative. Three studies \cite{cannavo2023immersive,guo2020improve,rothe2020reduce} indicated that partial or complete control over the participants’ FOV rotation can lead to noticeable motion sickness symptoms in the viewers, thereby negatively impacting their experience and sense of immersion/presence. However, in the experiment of Gugenheimer et al. \cite{gugenheimer2016swivrchair}, the research group designed an engineering device to control the swivel chair in which the participants were seated. This method effectively aligned physical sensations with virtual rotation, reducing motion sickness, and thus, the author and some other scholars thought this approach could yield better results \cite{gugenheimer2016swivrchair,rothe2019guidance}.

\subsubsection{The results of Robust variance estimation (RVE) methods}

Here, we report the estimated average effect sizes by the type of viewing modality, along with variance component estimates for each working model. Table~\ref{tab:coeffcient} contains all the results based on the correlated hierarchical effects (CHE) working model. With sampling correlations of $\rho = 0.6$, the estimated effects range from $0.961$ (SE = 0.294) for viewing modality with an agency to $0.068$ (SE = 0.328) for explicit diegetic guidance. Similar patterns were observed across different assumed sampling correlations ( $\rho$ = 0, 0.3, and 0.9), with effect sizes remaining consistent in both magnitude and direction. Based on robust Wald tests, we consistently failed to reject the null hypothesis that average effects are the same across viewing modalities (as $\rho = 0.6$, $p = 0.736$), and this finding remained stable across all assumed sampling correlations.

There are a few causes that might have contributed to this result. For instance, Participants’ previous VR experience was often considered a factor that could influence the final results. Among the 45 articles, only 13 (28.9\%) reported on participants’ VR experience. Most of these studies used simple statements to indicate participants’ familiarity with VR, such as “never” or “have experience” \cite{stebbins2019redirecting,aitamurto2021fomo}. Additionally, some studies \cite{vosmeer2017you} assessed participants’ VR experience using self-rated scales.

\subsection{Systematic Review: Quantitative Evaluations of User Experience}
Our statistical review of all the included CVR measurements shows that \emph{\textbf{"RQ2: What are the primary methods and key metrics currently used for measuring user experience in CVR?"}} remains an unanswerable question in the field. We concluded that there seemed not to be an established standard. However, using a questionnaire seemed to be a commonly accepted approach for evaluation. Further discussion about the lack of an established standard can be found in 5.2.

We organized and reported all the subjective measurements and objective data collection methods applied in all the experiments. Most studies only conducted subjective measurement after experiments (n=23), while some chose to apply objective measurements (n=9), such as eye or head tracking, scan-path, electrodermal activity (EDA) and dynamic depth of gaze (DDG), which were always carried out during the experience. Other studies (n=12) used both subjective and objective measurements.\\
\subsubsection{Questionnaires}
The most frequently used measurement was the questionnaire-based approach (33 of 45 papers, 73.3\%). Various validated and self-designed questionnaires were used for measuring user experience. Through our statistics, the most commonly measured dimensions of user experience were presence, immersion, narrative engagement, and motion sickness (Table \ref{tab:freq}).

\begin{itemize}
\item {\textbf{Presence}} 
\newline Of the 45 included papers, 42.2\% (n=19) used the questionnaire-based method to measure the sense of presence. The use of validated questionnaires included the iGroup Presence Questionnaire (iPQ; n=6, 31.6\%; \cite{schubert2001experience}), the Presence Questionnaire (PQ; n=2, 10.5\%; \cite{witmer1998measuring}), the Slater–Usoh–Steed Presence Questionnaire (n=2, 10.5\%; \cite{slater1994depth,usoh2000using}, and the ITC-SOPI questionnaire (n=2, 10.5\%; \cite{lessiter2001cross}). Some scales were used just a single time \cite{fox2009virtual,ahn2013effect,men2017impact}. Additionally, self-designed questionnaires or items were used in four studies (21.1\%).
\item {\textbf{Immersion}} 
\newline There was no significant preference shown in the questionnaires’ use of immersion, with seven papers in total measuring the sense of immersion (15.6\%). Some used the Immersion Experience Questionnaire (IEQ; n=2, 28.6\%; \cite{jennett2008measuring}) or adapted Roth’s Evaluation of Interactive Digital Narrative (IDN) toolbox \cite{roth2016evaluating} as the basis of the questionnaires (n=2, 28.6\%). Moreover, two studies (28.6\%) employed self-designed questionnaires or items (28.6\%).
\item {\textbf{Narrative Engagement}} 
\newline Another subjective metric that was always measured was narrative engagement (NE) (n=7, 15.6\%), which was highly correlated with CVR content. The most recognized and a relatively comprehensive questionnaire was the Measuring Narrative Engagement Questionnaire (MNEQ) (Busselle, \& Bilandzic, 2009). In our review, there were four references (57.1\%) that applied the MNEQ or adapted its items. The Narrative Transportation Scale (NTS, \cite{green2000role}) was used just once (14.3\%). Additionally, two studies (28.6\%) used self-designed items.
\item {\textbf{Motion Sickness}} 
\newline Of the 45 papers, 10 (22.2\%) measured motion sickness. The Simulator Sickness Questionnaire (SSQ) is widely acknowledged as the industry standard and commonly employed in the field \cite{kennedy1993simulator}. The SSQ was also the most frequently used questionnaire in our study (n=6, 60\%). In addition, one study used the Motion Sickness Assessment Questionnaire (MSAQ, \cite{gianaros2001questionnaire}) and three used self-designed questionnaires or items (37.5\%). 
\item {\textbf{Others}} 
\newline There were other metrics, depending on the research expectations. First, some papers measured usability or user experience (n=7, 15.6\%) using different trivial scales, such as the System Usability Scale \cite{bangor2008empirical} and the User Engagement Survey Short Form (UES-SF) \cite{o2018practical}. Second, some papers evaluated memory (n=6, 13.3\%), enjoyment (n=5, 11.1\%), empathy or affect (n=4, 8.9\%), cognitive load (n=3, 6.7\%), and attention \cite{vosmeer2017you}. Lastly, self-designed questionnaires often included items highly related to narrative content details. Researchers frequently used a straightforward and simple question to inquire about participants’ preferences for different experimental conditions.
\end{itemize}

\begin{table}[htb]
  \caption{The most common questionnaires on CVR.}
  \label{tab:freq}
  \begin{tabular}{llllcc}
    \toprule
    Name&Years&Authors&Concept/Term&Citations*&Items\\
    \midrule
    \ iPQ&2001&Schubert, et al. \cite{schubert2001experience}&Presence&2138&14\\
   \ PQ&1998&Witmer \& Singer \cite{witmer1998measuring}&Presence&7679&32\\
   \ SUS&1994/2000&Slater, et al. \cite{slater1994depth}/ Usoh, et al.\cite{usoh2000using}&Presence&1115/1736&3/6\\
   \ IEQ&2008&Jennett, et al.\cite{jennett2008measuring}&Immersion&2413&31\\
   \ MNEQ&2009&Busselle \& Bilandzic \cite{busselle2009measuring}&Narrative Engagement&1429&12\\
   \ SSQ&1993&Kennedy, et al. \cite{kennedy1993simulator}&Motion Sickness&5871&16\\
  \bottomrule
\end{tabular}
*Determined using Google Scholar,  December 2023
\end{table}
\subsubsection{Objective Measurements}
\begin{itemize}
    \item \textbf{Eye and head tracking}
    \newline The main objective evaluations involved eye and head-tracking techniques, since these would provide detailed data about users’ visual attention and perception which could show results such as attention time (AT) \cite{bender2019headset}. Furthermore, data collection is relatively easy with these techniques because HMDs are typically equipped with default eye and head trackers. In our review, 28.9\% of the papers (n=13) adopted this method. The results were often presented in the form of heatmaps \cite{rothe2018gazerecall,maranes2020exploring} or scan-path graphs \cite{knorr2018director,masia2021influence}.
\item \textbf{Others}
\newline Other objective measurement methods were also observed in this review, but their usage was not found to be significant. These methods involved the collection of physiological signal data from viewers, including the use of dilated capillary sclera \cite{knorr2018director}, DDG \cite{cao2020automatic}, and EDA \cite{curran2019understanding}.

\end{itemize}

\begin{table}[th]
\centering
\caption{Identical items in different questionnaires}
\renewcommand{\arraystretch}{1.5}
\scalebox{0.85}{
\begin{tabularx}{1.0\textwidth}{p{0.05\linewidth} X X X X X X}

\toprule
\textbf{Quest\newline ionna\newline ire} & \textbf{General presence/ Sense of being in VE} & \textbf{Awareness of the real world}& \textbf{Attention}& \textbf{Realism}& \textbf{Emotional \newline attachment}& \textbf{Sense of \newline control}\\
\midrule
iPQ \newline\cite{igroup} \newline \cite{schubert2001experience} & In the computer-generated world, I had a sense of “being there” &I was completely captivated by the virtual world. &How real did the virtual world seem to you?& & & \\
PQ \newline \cite{witmer1998measuring}& How much did the visual aspects of the environment involve you?& How aware were you of events occurring in the real world around you?& How well could you concentrate on the assigned tasks or required activities rather than on the mechanisms used to perform those tasks or activities?& To what degree did your experiences in the virtual environment seem consistent with your real-world experiences? & &To what degree did you feel that you were able to control events?\\
SUS \newline \cite{slater1994depth} \newline \cite{usoh2000using}& I had a sense of “being there” in the office space.& & &The extent to which, while immersed in the VE, it becomes more “real or present” than everyday reality.& & \\
IEQ \newline \cite{jennett2008measuring}& &To what extent was your sense of being in the game environment stronger than your sense of being in the real world? &To what extent did the game hold your attention? &To what extent did you feel that the game was something you were experiencing rather than something you were just doing? & To what extent did you feel emotionally attached to the game? & At any point did you find yourself becoming so involved that you were unaware you were even using controls?\\
MNEQ \newline \cite{kennedy1993simulator}& &At times during the program, the story world was closer to me than the real world.&I had a hard time keeping my mind on the program.& &The story affected me emotionally. & \\
\bottomrule
\end{tabularx}
}
\label{tab:auto_line_break}
\end{table}

\subsection{Different Terminologies Used in Questionnaires}
\label{subsec:4.4terminologies}
This section answers \emph{\textbf{"RQ3: What are the current issues with evaluation and metrics in measuring different viewing modalities?"}} We found that overlaps between the terms "presence," "immersion," and "narrative engagement" resulted in nonstandard questionnaire usage, with many shared subscales and items (refer to Table \ref{tab:auto_line_break}). This complicated the quantitative comparison of similar studies and raised concerns over the standardization of CVR measurements.

\section{DISCUSSION}
The goal of our meta-analysis and systematic review was to answer three research questions: \fixme{\emph{\textbf{RQ1, RQ2 and RQ3}}} (refer to Section~\ref{sec:introduction})\\
In our meta-analysis, \fixme{although we identified some potential impacts might be caused by certain modalities (see~\ref{subsec:4.2Meta}),} we observed notable variations in the reported effect sizes across different studies, even those with the same viewing modality. This inconsistency could be attributed to different experiment designs within the experiments. Through a detailed analysis, we identified the following potential factors: variables in experimental designs (see~\ref{subsec:5.1Variables}) and issues with CVR experience evaluation (see~\ref{subsec:5.2IssuesofCVR}). 

Researchers frequently used subjective measurements, especially questionnaires, with a focus on the concepts of presence or immersion.

One of the key challenges was the terminological confusion of “presence,” “immersion,” and “narrative engagement,” which were often used interchangeably in empirical practice. When evaluating the CVR experience, researchers encountered problems such as the irregular use of questionnaires, excessive use of self-developed questionnaires, and incomplete data reporting (see~\ref{subsec:5.2IssuesofCVR}).

\subsection{Variables in Experiment Design}
\label{subsec:5.1Variables}
In our exploration of CVR experimental designs, we observed that there were four main variables in experiment design across the 45 papers we examined—namely, stimulus, hardware, experiment design, and participants— even within the same modalities.

First, the choice of stimulus format and content might significantly have impacted viewer experience. In general, viewer immersion tended to increase in more realistic scenes and models \cite{cummings2016immersive}, but many studies often overlooked this factor. 

Second, equipment was another important variable. HMDs, including their resolution, refresh rate, and display technology (e.g., FOV size), directly impacted the user’s visual experience. Additionally, other devices, including the controllers mentioned in Section 4.3, could reduce viewers’ sense of presence. 

Third, the small sample sizes could have led to several potential problems \cite{checa2020review}, including inadequate statistical power (failing to detect actual differences or effects, even if they existed) and a higher risk of false negatives. Additionally, exposing the same participant to multiple treatments (for example, \cite{speicher2019exploring,cao2020automatic,bala2019elephant}) could have led to the user experiencing fatigue and learning aspects of the experiment after multiple exposures, while the effects from different treatments might have interacted with each other, thereby compromising the reliability of the results. 

Lastly, the selection of participants was also variable. Factors such as participants’ VR experience \cite{aitamurto2021fomo,husung2019portals,nielsen2016missing}and inclination toward relevant subject matter \cite{norouzi2021virtual,reyes2019combining} might directly have impacted experimental outcomes. Unfortunately, only a minority of the CVR studies took these factors into consideration. With a multitude of variables, it became challenging to compare results across different experiments.

\subsection{Issues of CVR Experience Evaluation}
\label{subsec:5.2IssuesofCVR}
In terms of evaluation, especially the questionnaires used, there were four main problems.

\emph{Irregular use of questionnaires. }In many cases, researchers do not use questionnaires properly. They may select several items from one questionnaire or questions from different questionnaires because there are no questionnaires specifically designed for assessing VR experiences or films (except for the MNEQ). For example, the most widely used iPQ questionnaire in CVR research is mainly (82.8\%) used for the mono monitor scenario \cite{igroup}, while the IEQ questionnaire was developed for measuring game experiences. The irregular use or poor design of the questionnaires can result in measurement errors \cite{muller2014survey,assila2016standardized}. 

\fixme{\emph{Over-reliance on SSQ.} Almost all CVR studies researching motion sickness used SSQ, a questionnaire that is not designed especially for VR scenarios. SSQ is widely regarded as an authoritative tool in the field of VR~\cite{Saredakis2020VRsickness,MacArthur2021Sick}, but its origins date back nearly three decades, raising questions about its continued relevance and suitability for new research in VR~\cite{Hirzle2021criticalSSQ}. Instead, many motion sickness questionnaires are more suitable for different CVR settings:  1) VRSQ \cite{KIM201866} is specially designed for VR scenarios, making it ideal for modern VR platforms and HMDs. Its limitation is its lack of items of non-oculomotor symptoms such as nausea~\cite{virtualworlds2010002}. 2) MSAQ \cite{gianaros2001questionnaire} is useable for studies focusing on multiple sensory and physical effects of motion, but it includes symptoms which may not be relevant for VR. 3) FMSQ \cite{FMSQ2011Behrang} is a one-item scale primarily intended for use in studies where motion sickness is considered a secondary issue. However, its simplicity makes it unsuitable for capturing the complexity and nuance of detailed symptom measurements. Overall, choosing a suitable motion sickness measurement tool is vital for ensuring that the study's findings are valid and trustworthy.}

\emph{Overlapping of different questionnaire content. }This issue is caused by identical items from different questionnaires (see~\ref{subsec:4.4terminologies}). Articles that use similar indicators often incorporate questionnaires with overlapping items, resulting in repetitive questions \cite{nilsson2016immersion,macquarrie2017cinematic}. This repetition can result in confusion among users when completing questionnaires.

\emph{Excessive use of self-designed questionnaires. }Just like other fields of HCI, self-designed questionnaires are widely used in CVR research, with 42.4\% (14 out of 33) either fully or partially utilizing self-designed questionnaires, due to the absence of standardized questionnaires \cite{diaz2021ux,diaz2019standardized}. These questionnaires are often tailored to match the specific stimulus of the experiment to assess “narrative accuracy” or “memory” \cite{godde2018cinematic,han2022evaluating,tong2022adaptive}. While they may be more suitable for the unique requirements of individual studies, their lack of reliability and validity makes it difficult to conduct parallel comparative studies.

\emph{Incomplete data reporting. }Some studies may have failed to compute or report the total or subscale scores of all items \cite{cao2019preliminary,maranes2020exploring,reyes2018measuring}, particularly when the data did not show significance. Additionally, some studies may have reported the mean (M) values without the standard deviation (SD) or just shown data with box plots or bar charts. These incomplete data reports can lead to biases in analyzing and drawing conclusions.

\subsection{Terminological Confusion and Research Implications}
\label{subsec:5.3ResearchImplications}
Terminological confusion is a significant issue in the CVR research field, notably exacerbating the inaccuracy of research results. On the one hand, terms such as “presence,” “immersion,” and “narrative engagement” overlapped between concepts, causing confusion among researchers \cite{nilsson2016immersion,macquarrie2017cinematic,mcmahan2013immersion}. On the other hand, many standardized questionnaires incorporated a number of identical items and subscales in their design (see~\ref{subsec:4.4terminologies}), further blurring the boundaries between different terms and worsening the terminological confusion phenomenon. Although different evaluations have different tendencies and motivations, IPQ mainly emphasizes the perception of VE, IEQ pays more attention to interactive engagement, and MNEQ pays more attention to narrative. However, in empirical research, the phenomenon of mixed or interchangeable use, such as using SUS presence questionnaires to measure sense of immersion \cite{speicher2019exploring} and combining different scales with different terms \cite{dining2017user,beck2021applying}, remains common. As a result, this makes it challenging to quantitatively compare the findings of similar studies, easily raising doubts regarding the standardization of CVR measurements.

\fixme{Furthermore, previous studies have not adequately explored the role of "attention," even though most viewing modalities (5 out of 6) are attention-driven, encompassing either visual guidance cues (Modalities 1-4) or physical rotation of the audience (Modality 6). None of these five modalities showed statistically significant effects in our meta-analysis. This may be because researchers often assumed that different modalities would generate varying levels of attention but did not confirm these assumptions through manipulation checks~\cite{speicher2019exploring, cao2020automatic, bala2019elephant}. For example, while Modalities 1-4 are classified by two attributes (explicit/implicit and diegetic/non-diegetic), there is no empirical evidence demonstrating their actual impact on attention levels. This lack of attention validation across modalities may have undermined the validity of previous findings.}

The insights from our systematic review and meta-analysis of the CVR field provide crucial guidance for producing more dependable results in future CSCW studies. CVR has proved to be a significant field in CSCW, specifically on topics such as social viewing in VR \cite{Siri2018Screen, Zamanifard2019Togetherness, Kukka2017SocialCity, montagud2022towards}, collaborative efforts in AR/VR for entertainment \cite{Long2023perspective,Zhang2021XRmas, kexuefu2023}, and the use of VR within work environments \cite{kohler2011avatar,Ortiz2023Workspace}. Ensuring consistent and rigorous experiment designs will lead to more reliable data. By addressing these methodological considerations, future CSCW research can yield more robust and generalizable findings, ultimately enhancing our understanding of how CVR technologies influence social interactions and collaborative engagement.

Overall, we suggest that researchers should aim for a better understanding of different definitions and terminologies in CVR within the HCI
field and endeavor to avoid the misuse or the interchangeable use of different terminologies. A taxonomy or a framework that clearly defines various terminologies and comprehensively delineates the essential aspects of CVR experience has the potential to alleviate terminological confusion in the field. \fixme{Moreover, we recommend that researchers should evaluate user’s attention level for each developed modality for better variable control.} In future user studies, researchers should employ questionnaires with greater care and rigor to produce dependable results. Additionally, investigating new standardized questionnaires designed specifically for CVR participants holds promise as a direction for further research.

\section{LIMITATIONS}
This study has limitations, which we hope to deal with in future work. First, it is possible that we may have missed some relevant papers in this review, despite our efforts to include the most pertinent keywords. Some papers may not have used these specific keywords in their titles or abstracts. Second, we mainly considered the sense of presence and immersion. As for other metrics, such as motion sickness and memory, we only conducted basic statistical analysis and did not include them in the meta-analysis. Furthermore, the participants of each study were usually of a very limited size (n=12–30), primarily due to the high cost associated with VR experiments. Consequently, this limitation may have led to less robust quantitative conclusions, since small sample sizes contribute to unstable effect size estimates. This means that the estimated effect sizes may have varied significantly between different studies, making it challenging to draw consistent conclusions. 

Moreover, the small number of effect sizes in each group severely constrained our ability to conduct meaningful moderator analyses. The limited data within each modality group meant that detecting significant moderators was extremely challenging, if not impossible. This limitation is particularly critical in a meta-analysis, where the identification of moderators can be crucial for understanding the nuances of effect size variability.

As a result, our study’s findings primarily hinge on the analysis of six separate meta-analyses, with the acknowledgment that the small sample sizes and the consequent inability to conduct comprehensive moderator analyses significantly limit the generalizability and depth of our conclusions.

\section{CONCLUSION}

In conclusion, our study provides a comprehensive review of the CVR field by conducting a systematic review and meta-analysis based on the data extraction and analysis of 45 articles. We primarily focused on exploring how different viewing modalities, including intervened rotation, avatar assistance, guidance cues, and perspective shifting,  influence CVR viewers’ experience. The study screened 3444 papers, selected 45 for systematic review and 13 of which also for meta-analysis. Although the \fixme{meta-analysis} results of previous studies showed inconsistency, \fixme{they also identified some potential impacts (positive and negative) of certain viewing modalities} and revealed additional issues within the CVR field. Specifically, terminological confusion and non-standardized evaluation methods may contribute to the lack of consistency in the results. Therefore, future research should address these issues to improve the rigor of CVR as a field in HCI.


\bibliographystyle{ACM-Reference-Format}
\bibliography{Reference_base}

\clearpage 

\section{Appendices}
\appendix

\begin{figure}[h]
\centering
\includegraphics[width=0.8\linewidth]{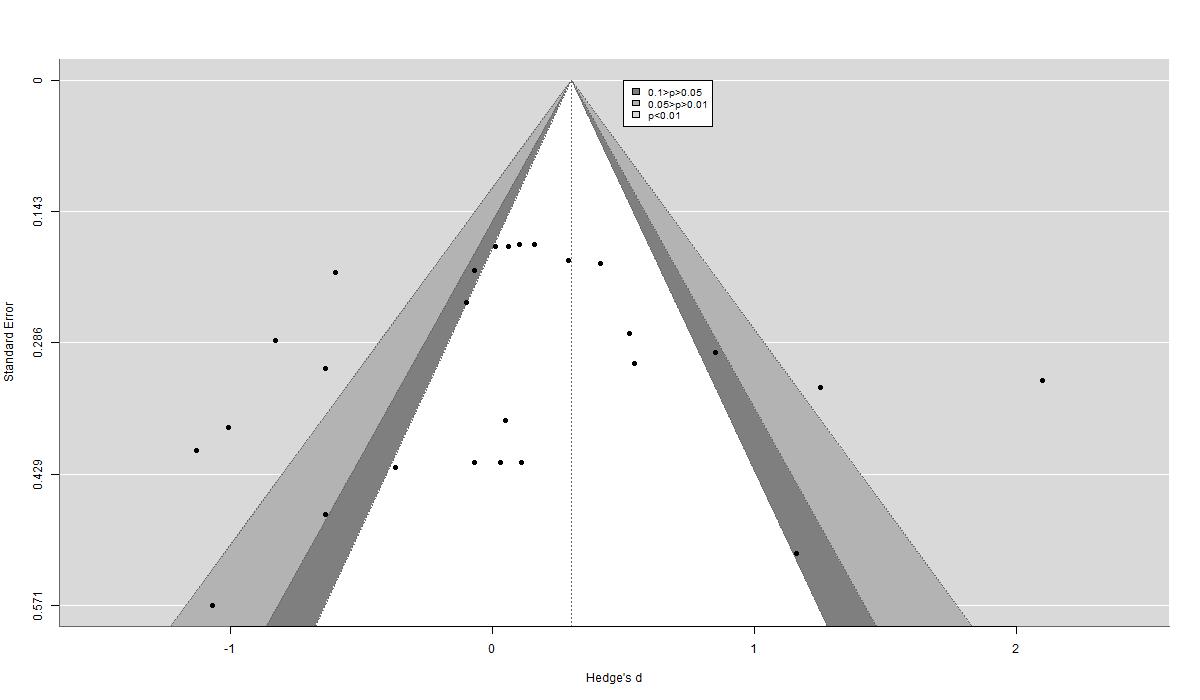}
\caption{Funnel plot}
\label{fig:appendix1}
\end{figure}

\newpage

\begin{figure}[H]
\centering
\rotatebox{90}{%
\begin{minipage}{\textheight}
\includegraphics[width=0.9\linewidth]{table5_Meta_Information.pdf}
\caption{Information of Studies Included in Meta-analysis}
\label{fig:appendix2}
\caption*{These two articles \cite{speicher2019exploring,cao2020automatic}reported multiple effect sizes. To maintain the independence of data related to the same viewing modality, when multiple effect sizes are reported for the same viewing modality involving the same participants, only one effect size is retained.}
\end{minipage}
}
\end{figure}

\begin{table}[th]
\centering
\caption{Details of questionnaires and terminologies usage}
\label{tab:Q&T}
\renewcommand{\arraystretch}{1.5}
\scalebox{0.85}{
\begin{tabularx}{1.1\textwidth}{m{0.03\linewidth} m{0.05\linewidth} p{0.1\linewidth} m{0.1\linewidth} X m{0.11\linewidth} X}
\toprule
\textbf{ES\newline ID} &\textbf{Study\newline ID} &\textbf{Author}& \textbf{Question-\newline naire}&\textbf{Questionnaire Details}& \textbf{Terminology Confusion}&\textbf{Terminology Details}\\
\midrule

1 &1 &\centering Beck\&Rothe, 2021\cite{beck2021applying}&\centering Adapted&4 items adapted from IPQ\cite{igroup}, \newline5 items adapted from IEQ\cite{jennett2008measuring} &\centering YES&Used "presence" and "immersion" interchangeably  \\

2 & 2& \centering Rothe et al., 2020\cite{rothe2020reduce}& \centering Adapted& Standerlized use of IPQ\cite{igroup}, and 5 items adapted from SSQ\cite{kennedy1993simulator}, and some self-developed questions &\centering NO &Used "presence"\\

3 & 3&\centering Guo et al., 2020\cite{guo2020improve} &\centering Valid&IEQ\cite{jennett2008measuring} &\centering NO& Used "immersion" \\

4&4 &\centering Chen et al., 2017\cite{chen2017effect} &\centering Valid&IPQ\cite{igroup} &\centering NO& Used "presence"\\

5\newline 6\newline 7\newline 8&5 &\centering Speicher et al., 2019\cite{speicher2019exploring}&\centering Selected items& A single question from SUS\cite{slater1994depth,usoh2000using}, 
and other questions from MASQ\cite{gianaros2001questionnaire}, UEQ\cite{schrepp2017design},  NASA-TLX\cite{HART1988139}.&\centering YES &Used "presence" and "immersion" interchangeably \\

9\newline 10&6 &\centering Nielsen et al., 2016\cite{nielsen2016missing} &\centering Adapted&3 items adapted from SUS \cite{slater1994depth,usoh2000using}  &\centering NO& Used "presence"\\

11\newline 12\newline 13&7 &\centering Cao et al., 2020\cite{cao2020automatic} &\centering Self-developed&Self-designed questionnaire \newline including 5 questions & \centering NO& Used "immersion"\\

14\newline 15\newline 16\newline 17&8 &\centering Bala et al., 2019\cite{bala2019elephant} &\centering Valid&Narrative Transportation Scale (NTS)\cite{green2000role}   & \centering NO& Used "immersion"\\

18\newline 19&9 &\centering Dining, 2017\cite{dining2017user} &\centering Adapted&Structure adapted from IPQ\cite{igroup}, and a general question from SUS  \cite{slater1994depth,usoh2000using}  & \centering YES& Used "presence" , "immersion", and "narrative engagement/ story presence" interchangeably\\

20\newline 21&10 &\centering Tong et al., 2022\cite{tong2022adaptive} &\centering Valid&Combine UES-SF, 12 items from MOOCs\cite{ip2018design}, and three-part evaluation from\cite{shah2020real}& \centering YES& Used "engagement", "presence" and "immersion" interchangeably\\

22&11&\centering Gugenheimer et al., 2016 \cite{gugenheimer2016swivrchair}&\centering Valid&Used $E^2$I  questionnaire\cite{Lin2002E2l}, \newline and RSSQ\cite{kim1999development}&\centering NO&Used "presence"\\

23&12&\centering Lin et al., 2017 \cite{lin2017tell}&\centering Adapted&Used valid SSQ \cite{kennedy1993simulator}and simple questions related feeling of presence and enjoyment.&\centering YES&Used "presence" and "engagement" interchangebaly\\

24&13&\centering Norouzi et al., 2021 \cite{norouzi2021virtual}&\centering Valid&Used SUS\cite{slater1994depth,usoh2000using}, UEQ\cite{schrepp2017design}, fear of missing out questions\cite{macquarrie2017cinematic}, Preference Questionnaire\cite{wallgrun2020comparison}, and SSQ\cite{kennedy1993simulator}&\centering NO&Used "presence"\\
\bottomrule
\end{tabularx}
}
\end{table}

\begin{table}[htb]
\centering
\small
\caption{Coefficient Studies and Effect Sizes}
\label{tab:coeffcient}
\begin{tabular}[\linewidth]{lcccccccccc}
\toprule
& & & \multicolumn{4}{c}{CHE Estimated effects (SE)} \\
\cmidrule{4-7}
Coefficient & Studies & Effect sizes & $\rho=0$ & $\rho=0.3$ & $\rho=0.6$ & $\rho=0.9$ \\
\midrule
\multicolumn{7}{l}{\textbf{Modality}} \\
With explicit diegetic guidance & 3 & 3 & 0.130 & 0.099 & 0.069 & 0.04\\
&&&(0.336) & (0.332) & (0.328) & (0.325) \\
With explicit non-diegetic guidance & 7 & 7 & 0.460 & 0.467 & 0.475 & 0.485 \\&&&(0.441) &(0.448) &(0.455) & (0.463) \\
With implicit non-diegetic guidance & 6 & 6 & 0.255 & 0.230 & 0.205 & 0.180\\
&&& (0.245) & (0.258) & (0.272) & (0.285) \\
With agency & 2 & 2 & 0.344 & 0.325 & 0.309 & 0.295 \\
&&& (0.342) & (0.341) & (0.342) & (0.344) \\
Limited rotation & 2 & 2 & 0.991 & 0.975 & 0.961 & 0.950 \\
&&& (0.284) & (0.288) & (0.294) & (0.299) \\
Forced rotation & 6 & 6 & 0.360 & 0.346 & 0.334 & 0.321\\
&&& (0.636) & (0.636) & (0.637) & (0.638) \\
Wald test p value & & & 0.746 & 0.740 & 0.736 & 0.735 \\
\midrule
\multicolumn{7}{l}{\textbf{Questionnaire}} \\
Valid (reference group) & 5 & 7 \\
Adapted & 1 & 4 & -0.196 & -0.182 & -0.169 & -0.160 \\
&&& (0.443) & (0.451) & (0.460) & (0.468) \\
Selected\_items & 1 & 3 & -0.189 & -0.168 & -0.146 & -0.127 \\
&&& (0.333) & (0.330) & (0.329) & (0.328) \\
Self\_developed & 8 & 12 & 0.079 & 0.089 & 0.100 & 0.111 \\
&&& (0.179) & (0.168) & (0.192) & (0.199) \\
Wald test p value & & & 0.926 & 0.933 & 0.935 & 0.930 \\
\midrule
\multicolumn{7}{l}{\textbf{Terminology}} \\
Not mixed (reference group) & 8 & 13 \\
Mixed & 6 & 13 & -0.530 & -0.521 & -0.513 & -0.505 \\
&&& (0.328) & (0.332) & (0.337) & (0.342) \\
Wald test p value & & & 0.189 & 0.199 & 0.209 & 0.219\\
\midrule
$\hat{T}$ & & & 0 & 0 & 0 & 0 \\
$\hat{\omega}$ & & & 0.590 & 0.617 & 0.649 & 0.685\\
\bottomrule
\end{tabular}
\end{table}

\end{document}